\documentclass[12pt,a4paper]{article}
\usepackage{amsmath,amssymb}
\usepackage{mathtools}
\usepackage{bbold}
\usepackage{graphicx}
\usepackage[colon,authoryear]{natbib}
\usepackage[ruled]{algorithm2e}
\usepackage{listings}
\usepackage{enumitem}
\usepackage{hyperref}
\usepackage[misc,geometry]{ifsym}
\usepackage{longtable}
\usepackage{siunitx}       
\usepackage{tikz,tkz-graph}
\usepackage[customcolors]{hf-tikz}
\usepackage{authblk}

\tikzset{style green/.style={
    set fill color=green!50!lime!60,
    set border color=white,
  },
  style cyan/.style={
    set fill color=cyan!90!blue!60,
    set border color=white,
  },
  style orange/.style={
    set fill color=orange!80!red!60,
    set border color=white,
  },
  hor/.style={
    above left offset={-0.03,0.31},
    below right offset={0.03,-0.125},
    #1
  },
  ver/.style={
    above left offset={-0.03,0.3},
    below right offset={0.03,-0.15},
    #1
  }
}

\def\BigColSep{\setlength{\arraycolsep}{3pt}}

\hypersetup{
 colorlinks=true,
 citecolor=blue,
}

\begin{document}

\title{Best of both worlds?\thanks{This work was supported by the University of Malaya High Impact Research Grant UM.C/625/1/HIR/MOHE/SC/13.}\\
\large Simultaneous evaluation of researchers and their works
}

\author[1]{Ephrance Abu Ujum}
\author[2]{Gangan Prathap}
\author[1]{Kuru Ratnavelu}
\affil[1]{\footnotesize Institute of Mathematical Sciences, University of Malaya, 50603 Kuala Lumpur, Malaysia \\
\url{ephrance@siswa.um.edu.my},~\url{kuru@um.edu.my}
}
\affil[2]{\footnotesize CSIR National Institute for Interdisciplinary Science and Technology (NIIST), Council of Scientific and Industrial Research, Thiruvananthapuram, 695 019, Kerala, India\\
\url{gp@niist.res.in}
}

\maketitle
\begin{abstract}
This paper explores a dual score system that simultaneously evaluates the relative importance of researchers and their works.
It is a modification of the CITEX algorithm recently described in~\cite{2015arXiv150104894P}.
Using available publication data for $m$ author keywords (as a proxy for researchers) and $n$ papers it is possible to construct a $m \times n$ author-paper feature matrix.
This is further combined with citation data to construct a HITS-like algorithm that iteratively satisfies two criteria:
first, \emph{a good author is cited by good authors},
and second, \emph{a good paper is cited by good authors}.
Following Pal and Ruj, the resulting algorithm produces an author eigenscore and a paper eigenscore.
The algorithm is tested on 213,530 citable publications listed under Thomson ISI's ``\emph{Information Science \& Library Science}'' JCR category from 1980--2012.

\end{abstract}

\section{Introduction}
\label{sec:caps-intro}
Rankings provide an effective means to artificially assign order to the ever increasing volume of published research and researchers.
The study and development of such work is increasingly trending towards what could be termed as \emph{bibliometric analytics},
which we define here as\footnote{We note that~\cite{bhatt2009topics} and~\cite{rethlefsen2013environmental} used the term ``\emph{bibliometric analytics}'' but have not provided a formal definition.} 
``key indicators derived from bibliometric data through mathematical or statistical analysis for the purpose of generating insight''.
In addition to information retrieval, bibliometric analytics focuses on discovering patterns specific to the data at hand in order to support decision-making or inference-related tasks.
This paper is yet another step in this direction.

Specifically, this paper builds on recent work established by~\cite{2015arXiv150104894P} to simultaneously score research authors and papers by relative importance. 
The proposed algorithm, dubbed CITEX (CITation indEX), takes advantage of the many-to-many correspondence between a given set of authors and the papers they have collectively published.
For this purpose, mappings between both sets can be formalized as linkages on a bipartite graph, hereon referred to as the author-paper (or author-document) network.
In this sense, the cumulative advantage accrued by authors due to their papers and vice versa can be quantified using graph theoretic methods.

Furthermore, papers are interconnected through citation links; 
that is, a typical paper refers to previous works in order to acknowledge \emph{relevance}\footnote{In general, citation linkages are made to indicate \emph{reaction} to past work rather than concrete \emph{dependence}.
Hence, the presence of citation linkages -- that is, a link pointing from citing (referring) paper to cited (referred) paper -- serves to describe intellectual flows in successive works, which in itself does not necessarily imply a flow of influence. 
} in addition to specifying its own \emph{placement}\footnote{This is a notion of the paper's \emph{location} as opposed to its~\emph{position}.
As with a citation count, the presence of a citation link does not explicitly convey whether it takes on the position of supporting or opposing the referred work.
} within the existing literature.
The resulting paper citation network can thus be represented as a directed graph.
Since the distribution of citation links varies from one paper to the next -- usually in a highly skewed manner~\citep{simon1955class,de1965networks,price1976general,newman2009first} -- this can then be used as a basis to distinguish which papers are more prominently located than others.
Several schemes have been proposed to exploit precisely this feature; 
i.e.~scores are computed for each paper based on some discriminatory function of its connectivity (or how it is embedded within a structure of links)~\citep{chen2007finding}.
These papers can then be ordered according to the computed scores to produce rankings.
Such schemes are integral to information retrieval tasks on online databases, for example, Google Scholar, CiteSeerX, and Microsoft Academic Search.

CITEX extends this tradition by combining information from the author-paper network with the paper citation network to determine which authors and which papers stands out more than others. 
The development of such algorithms are important to explore alternative means of assembling bibliometric indicators (and their derived rankings) through purposeful integration of available information.
CITEX is interesting in its construction because it provides a coupled dual score system: a relative importance score for authors and another for papers, 
hence, the relative standings of knowledge creators and the results of their labours can be determined within a single framework.
Simply put, CITEX asserts that: (1) good authors are either highly prolific with, or are highly cited by good authors; and, 
(2) good papers share the same authors with, or are cited by good papers.

This paper is organized as follows.
We provide an in-depth discussion on the construction of the CITEX algorithm in Section~\ref{sec:citex-algo-construction} and a critique is offered in Section~\ref{sec:citex-algo-inspection}.
Our proposed modification, hereon referred to as the CAPS (Coupled Author-Paper Scoring) algorithm, is then described in detail in Section~\ref{sec:caps-method}.
To provide a point of comparison, both algorithms are tested on a real world dataset in Section~\ref{sec:caps-data}.
This consists of 200,000+ ISI-cited papers published from 1980-2012 listed under the Journal Citation Reports subject category of ``\emph{Information Science \& Library Science}''. 
The paper is concluded in Section~\ref{sec:caps-conclusion}.

\section{The CITEX algorithm}
\label{sec:citex-algo-construction}
Suppose we are presented with a corpus consisting of $m$ authors and $n$ papers.
Furthermore, suppose that from this corpus, we are able to extract the binary $m \times n$ author-paper feature matrix, $M$, and binary $n \times n$ citation matrix, $C$.
Let an entry $M_{ij} = 1$ denote that author $i$ on the $i$-th row of $M$ has (co)authored paper $j$ on the $j$-th column of $M$ ($M_{ij} = 0$ otherwise).
This implies that row sums of $M$ correspond to total papers published by each author.
Column sums of $M$ correspond to total authors for each paper.
A column-normalized version of $M$ (with the same dimensions) can be constructed so that \emph{authorship share} of author $i$ to paper $j$ is divided equally as $W_{ij} = M_{ij}/{\sum_i M_{ij}}$.

In a similar way, let $C_{ij} = 1$ denote that cited paper $j$ on the $j$-th column of $C$ receives a citation from a citing paper $i$ on the $i$-th row of $C$ ($C_{ij} = 0$ otherwise).
Additionally, we require that $C$ contains no self-citations ($C_{ii} = 0$).
Given an extreme case where $C = \mathbb{0}_{n \times n}$, Pal and Ruj define the CITEX paper and author scores as $y_j = \sum_{i=1}^m M_{ij} x_i$ and $x_i = \sum_{j=1}^n W_{ij} y_j$, respectively.
These expressions are written in matrix form as $y \leftarrow M^T x$ and $x \leftarrow W y$.
This captures the notion that the $y$-score for paper $j$ depends on the relative importance of its authors, while the $x$-score for author $i$ depends on her authorship share ($W_{ij}$) for each paper $j$ multiplied by its corresponding score $y_j$.

A complete description however requires the inclusion of citation features.
Since this must reduce to the case of a zero citation matrix, Pal and Ruj achieve this by the inclusion of a $(I + C^T)$ term (which is equivalent to adding in paper self-citations to $C$).
Since $y \leftarrow M^T W y$ and $x \leftarrow W M^T x$, then for the $k$-th recursion:
\begin{align}
	x^{(k)} &= W(I + C^{T})M^{T} x^{(k-1)} \label{eqn:citex-rule-1} \\
	y^{(k)} &= (I + C^{T})M^{T}W y^{(k-1)} \label{eqn:citex-rule-2}
\end{align}
is one such possible choice.
By induction, we obtain:
\begin{align}
	x^{(k)} &= [W(I + C^{T})M^{T}]^k x^{(0)} \\
	y^{(k)} &= [(I + C^{T})M^{T}W]^k y^{(0)}
\end{align}
For initial guess vectors, Pal and Ruj use $x^{(0)} = \mathbb{1}_{m \times 1}$ and $y^{(0)} = \mathbb{1}_{n \times 1}$.
Supposing $P = W(I + C^T)M^T$, so that $x^{(k)} = P^k x^{(0)}$, then:
\begin{equation}
x^{(k + 1)} = P P^k x^{(0)} = P x^{(k)} \label{eqn:citex-authorscore-recursion}
\end{equation} 
If the distance between two $x$ score vectors is $\| x^{(k+1)} - x^{(k)} \| < \epsilon$ then convergence is met relative to tolerance $\epsilon$~\citep{franceschet2011pagerank}.
Since $P$ is a nonnegative matrix with dimensions $n \times n$ and $x^{(0)} > 0$, then in accordance with the Perron-Frobenius theorem\footnote{In particular, given that $P x = c x$ and $c = 1$ is the largest eigenvalue, then $P^k x^{(0)}$ converge to a vector $x^{*}$ (in the same direction as $x$) as $k \rightarrow \infty$.}, the $x$ scores become stationary as $k \rightarrow \infty$, thus satisfying $P x^{*} = x^{*}$~\citep{perron1907theorie,frobenius1912matrizen}.
A similar argument is applicable for $y$ by setting $Q = (I + C^T) M^T W$.

There are other algorithms that combine author and paper features.
One notable example is the Co-Ranking framework proposed in~\cite{zhou2007co}.
This approach uses a PageRank-based model on a bipartite co-authorship/paper citation network, whereby two intra-class random walks allow traversal strictly between one class of nodes, while an inter-class random walk allows jumps between networks.
The stationary probabilities for author nodes and paper nodes are computed by coupling the random walks (assuming the status of researchers and the work they produce are mutually reinforced). 
The resulting algorithm yields improvements compared to when applying PageRank on either feature (network) in isolation, although at the expense of introducing three additional adjustable parameters to the usual one-parameter PageRank\footnote{We are referring to the \emph{damping parameter} originally described in~\cite{brin1998anatomy}.
The interested reader is referred to~\cite{langville2006google} and~\cite{chen2007finding} for an in-depth discussion on the PageRank algorithm.
}.
CITEX adds an interesting twist to the current literature since, unlike PageRank, it does not depend on any adjustable parameters.

\section{Expected behaviour and blindspots}
\label{sec:citex-algo-inspection}
Since the performance of a data mining algorithm depends on its design~\citep{jahne2000computer,balakin2009pharmaceutical},
it is useful to determine precisely what features are emphasized by CITEX in order to anticipate the qualitative aspects of the ranking it will necessarily produce.
In particular, we are interested in the conditions that maximize a given score since the highest percentile is designed to correspond to the topmost ranks.
Specific to the CITEX author score, Equation~\ref{eqn:citex-rule-1} can be expanded as:
\begin{align}
	x^{(k)} &= W M^T x^{(k-1)} + W C^T M^T x^{(k-1)} \label{eqn:citex-rule-1-expanded} \\
	x^{(k)}_i &= \sum_{a=1}^m \sum_{p=1}^n W_{i p} M_{a p} x^{(k-1)}_a
		+ \sum_{a=1}^m \sum_{p_1,p_2=1}^n W_{i p_1} C_{p_2 p_1} M_{a p_2} x^{(k-1)}_a \label{eqn:citex-rule-1-summation-form}
\end{align}
The first term on the right hand side of Equation~\ref{eqn:citex-rule-1-summation-form} captures the \emph{cumulative authorship share} of author $i$ with author $a$.
This term is positively biased towards author $i$ if she is prolific (adjusting for authorship share), and more so if she collaborates frequently with ``good authors'' (those with high $x$-scores).
This includes the case where $a=i$, so that if the cumulative authorship share of $i$ herself is significantly large, then $x_i^{(k)} \sim x_i^{(k-1)} \sum_{p=1}^n W_{ip}$.

As for the second term, a citation from paper $p_2 \rightarrow p_1$ corresponds to an author citation from $a \rightarrow i$ fractionalized by $W_{ip_1}$.
Hence, this term increases the larger the number of citations from $a \rightarrow i$, the larger the authorship share for each paper authored by $i$ (for which credit is minimally split), and the larger the $x$-score of $i$'s citing authors.
Put together, \emph{CITEX defines a good author as one who publishes frequently with good authors, and is even more so if he/she is cited by good authors}.

A similar analysis can be done for the CITEX paper score as given in Equation~\ref{eqn:citex-rule-2}:
\begin{align}
	y^{(k)} &= M^T W y^{(k-1)} + C^T M^T W y^{(k-1)} \label{eqn:citex-rule-2-expanded} \\
	y^{(k)}_j &= \sum_{a=1}^m \sum_{p=1}^n M_{a j} W_{a p} y^{(k-1)}_p
		+ \sum_{a_1,a_2=1}^m \sum_{p=1}^n C_{p j} M_{a_1 j} W_{a_2 p} y^{(k-1)}_p \label{eqn:citex-rule-2-summation-form}
\end{align}
From the right hand side of Equation~\ref{eqn:citex-rule-2-summation-form}, we see again that CITEX defines relative importance in terms of two components; the first term captures publication features while the second term captures citation features.

For the first term, we see that paper $j$ receives fractional $y$-score contributions for each author $a$ appearing in both papers $j$ and $p$.
Essentially, $M_{aj} W_{ap}$ is an \emph{author similarity} term, hence, this part of the equation increases for papers that share the same authors.
This term will also increase the larger the $y$-score for each ``similar author'' paper $p$ (relative to $j$) and whenever $W_{ap} \rightarrow 1$.
For the case of an author $i$ with a significantly large number of papers, we could end up with $y^{(k)}_j \sim y^{(k-1)}_j \sum_{p=1}^n W_{ip}$. 

For the second term, we see that $y_j$ depends on the sum of $y$-scores from each paper that cites it, $p$ (``good papers'' have high $y$-score). 
With some rearranging, the second term also contains the product $W_{a_2 p} C_{pj} M_{a_1 j}$.
This means that the $y$-score of paper $j$ depends on the sum of fractionalized citations from all citing papers $p$ (i.e. $\sum_{p=1}^n W_{a_2 p}C_{pj}$).
Combining this with the effect from the first term of Equation~\ref{eqn:citex-rule-2-summation-form}, we surmise that \emph{CITEX defines a good paper as one with high author similarity with good papers, and is even more so if it is cited by good papers}.

\begin{figure}[hbt!]
\begin{align*}
M &=
\begingroup\BigColSep 
\begin{bmatrix}
1 & 1 & 1 & 0 & 0 & 0 \\
0 & 0 & 0 & 1 & 1 & 0 \\
0 & 0 & 0 & 0 & 0 & 1 \\
\end{bmatrix}
\endgroup
&
\begin{tikzpicture}[baseline=-20pt]
	\tikzset{VertexStyle/.style={shape=circle,red,minimum size=0.1pt,fill}}
	\Vertex[x=0,y=0.0,LabelOut=true,Lpos=180]{$a_1$}
	\Vertex[x=0,y=-0.6,LabelOut=true,Lpos=180]{$a_2$}
	\Vertex[x=0,y=-1.2,LabelOut=true,Lpos=180]{$a_3$}
	\tikzset{VertexStyle/.style={shape=rectangle,blue,minimum size=0.5pt,fill}}
	\Vertex[x=3,y=0.4,LabelOut=true,Lpos=0,Ldist=10pt]{$p_1$}
	\Vertex[x=3,y=0.0,LabelOut=true,Lpos=0,Ldist=10pt]{$p_2$}
	\Vertex[x=3,y=-0.4,LabelOut=true,Lpos=0,Ldist=10pt]{$p_3$}
	\Vertex[x=3,y=-0.8,LabelOut=true,Lpos=0,Ldist=10pt]{$p_4$}
	\Vertex[x=3,y=-1.2,LabelOut=true,Lpos=0,Ldist=10pt]{$p_5$}
	\Vertex[x=3,y=-1.6,LabelOut=true,Lpos=0,Ldist=10pt]{$p_6$}
	\Edges[lw=0.5pt]($a_1$,$p_1$)
	\Edges[lw=0.5pt]($a_1$,$p_2$)
	\Edges[lw=0.5pt]($a_1$,$p_3$)
	\Edges[lw=0.5pt]($a_2$,$p_4$)
	\Edges[lw=0.5pt]($a_2$,$p_5$)
	\Edges[lw=0.5pt]($a_3$,$p_6$)
	\tikzset{EdgeStyle/.style={->,bend left}}
	\Edges($p_4$,$p_5$)
	\Edges($p_4$,$p_6$)
	\Edges($p_5$,$p_6$)
\end{tikzpicture}
&
C =
\begingroup\BigColSep 
\begin{bmatrix}
0 & 0 & 0 & 0 & 0 & 0 \\
0 & 0 & 0 & 0 & 0 & 0 \\
0 & 0 & 0 & 0 & 0 & 0 \\
0 & 0 & 0 & 0 & 1 & 1 \\
0 & 0 & 0 & 0 & 0 & 1 \\
0 & 0 & 0 & 0 & 0 & 0 
\end{bmatrix}
\endgroup
\end{align*}
\caption{Problem 1 --- Hypothetical case of one prolific solo author with no citations.
CITEX gives $x = [0.333, 0.333, 0.333]$ and $y = [0.143, 0.143, 0.143, 0.095, 0.191, 0.285]$.
Hence, all three authors are ranked equally even though there are stark qualitative differences between their publication and citation patterns.
Understandably, paper $p_6$ has the highest score followed by $p_5$ due to the number of citations they receive compared to no citations for the other papers.
Oddly, $p_4$ is ranked lower than papers $p_1$, $p_2$ and $p_3$ despite being authored by author $a_2$ who has one citation more than $a_1$ (via $p_5$).
}
\label{fig:caps-toy-model-1}
\end{figure}
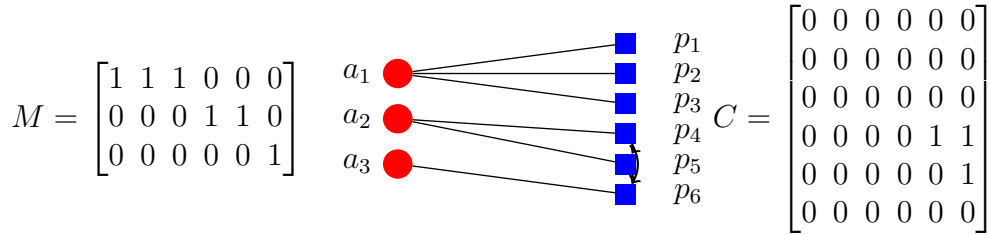

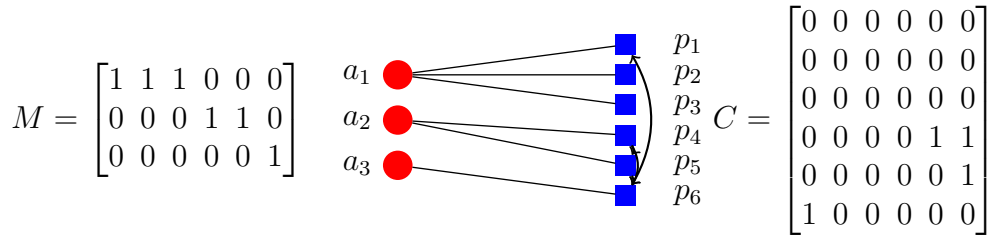
\begin{figure}[hbt!]
\begin{align*}
M &=
\begingroup\BigColSep 
\begin{bmatrix}
1 & 1 & 1 & 0 & 0 & 0 \\
0 & 0 & 0 & 1 & 1 & 0 \\
0 & 0 & 0 & 0 & 0 & 1 \\
\end{bmatrix}
\endgroup
&
\begin{tikzpicture}[baseline=-20pt]
	\tikzset{VertexStyle/.style={shape=circle,red,minimum size=0.1pt,fill}}
	\Vertex[x=0,y=0.0,LabelOut=true,Lpos=180]{$a_1$}
	\Vertex[x=0,y=-0.6,LabelOut=true,Lpos=180]{$a_2$}
	\Vertex[x=0,y=-1.2,LabelOut=true,Lpos=180]{$a_3$}
	\tikzset{VertexStyle/.style={shape=rectangle,blue,minimum size=0.5pt,fill}}
	\Vertex[x=3,y=0.4,LabelOut=true,Lpos=0,Ldist=10pt]{$p_1$}
	\Vertex[x=3,y=0.0,LabelOut=true,Lpos=0,Ldist=10pt]{$p_2$}
	\Vertex[x=3,y=-0.4,LabelOut=true,Lpos=0,Ldist=10pt]{$p_3$}
	\Vertex[x=3,y=-0.8,LabelOut=true,Lpos=0,Ldist=10pt]{$p_4$}
	\Vertex[x=3,y=-1.2,LabelOut=true,Lpos=0,Ldist=10pt]{$p_5$}
	\Vertex[x=3,y=-1.6,LabelOut=true,Lpos=0,Ldist=10pt]{$p_6$}
	\Edges[lw=0.5pt]($a_1$,$p_1$)
	\Edges[lw=0.5pt]($a_1$,$p_2$)
	\Edges[lw=0.5pt]($a_1$,$p_3$)
	\Edges[lw=0.5pt]($a_2$,$p_4$)
	\Edges[lw=0.5pt]($a_2$,$p_5$)
	\Edges[lw=0.5pt]($a_3$,$p_6$)
	\tikzset{EdgeStyle/.style={->,bend left}}
	\Edges($p_4$,$p_5$)
	\Edges($p_4$,$p_6$)
	\Edges($p_5$,$p_6$)
	\tikzset{EdgeStyle/.style={->,bend right}}
	\Edges($p_6$,$p_1$)
\end{tikzpicture}
&
C =
\begingroup\BigColSep 
\begin{bmatrix}
0 & 0 & 0 & 0 & 0 & 0 \\
0 & 0 & 0 & 0 & 0 & 0 \\
0 & 0 & 0 & 0 & 0 & 0 \\
0 & 0 & 0 & 0 & 1 & 1 \\
0 & 0 & 0 & 0 & 0 & 1 \\
1 & 0 & 0 & 0 & 0 & 0 
\end{bmatrix}
\endgroup
\end{align*}
\caption{Problem 2 --- The effect of high author similarity with good papers.
The setup in this diagram is similar to Figure~\ref{fig:caps-toy-model-1} with one additional citation link added from paper $p_6$ to $p_1$.
CITEX gives $x = [0.521, 0.214, 0.214]$ and $y = [0.243, 0.175, 0.175, 0.068, 0.136, 0.203]$.
Author $a_1$ leads by author score followed by a tie between $a_2$ and $a_3$, despite the absence  of (co)author self-citations to $a_1$ (note that $a_2$ has one author self-citation via $p_4 \rightarrow p_5$).
Paper $p_1$ is ranked highest despite having only one citation because it is cited by a good paper ($p_6$).
Due to the way paper scores are propagated in CITEX, papers $p_2$ and $p_3$ also receive high scores just by having high author similarity with paper $p_1$.
}
\label{fig:caps-toy-model-2}
\end{figure}

Based on our analysis, we have determined two quirks with the original formulation of CITEX.
These are:
\begin{enumerate}
	\item $x_i^{(k)} \sim x_i^{(k-1)} \sum_{p=1}^n W_{ip}$:
	the CITEX author score for an author $i$ can increase from being highly prolific, and more so if he/she tends to coauthor in small teams.
	This allows for the case of an extremely prolific solo author to be over-represented by the algorithm.
He or she may not even need a boost from citation count (from good authors or otherwise) in order to obtain a high CITEX author score.
	\item $y_j^{(k)} \sim y_j^{(k-1)} \sum_{p=1}^n W_{ip}$: 
	the CITEX paper score can increase just by having the same author list repeat over a significant fraction of the collection, with this effect becoming more pronounced if the listing tends to be short.
	Similarly, such cases can be over-represented by CITEX without a boost from citation count (from good papers or otherwise).
\end{enumerate}
To illustrate the potential problems associated with these quirks, we construct two toy calculations analogous to those posed in~\cite{2015arXiv150104894P}.
These are as shown in Figure~\ref{fig:caps-toy-model-1} and Figure~\ref{fig:caps-toy-model-2}.

As a result of the quirks highlighted in Figure~\ref{fig:caps-toy-model-1} and Figure~\ref{fig:caps-toy-model-2}, we can expect that author and paper rankings generated by CITEX will suffer from specificity issues since extreme publication and citation traits are mixed together.
The task of this paper is to propose a more elegant variation of the CITEX algorithm that addresses the above mentioned issues.

\section{An improved Coupled Author-Paper Scoring algorithm}
\label{sec:caps-method}
As highlighted in Section~\ref{sec:citex-algo-inspection}, CITEX has a built-in tendency to produce a rank ordering that gives undesired priority to highly productive authors (even if they are relatively uninfluential), in addition to assigning high relative importance to papers associated to highly prolific authors (overriding the citation impact of other papers). 

To circumvent these issues, we propose dropping the self-citation term $(I + C^T)$ in Equations~\ref{eqn:citex-rule-1} and~\ref{eqn:citex-rule-2}, and replace the $M$ matrices with $W$ matrices to ensure conservation of citation count when switching from the paper citation network to the author citation network (inter-author citations are fractionalized).
This results in the following set of equations which defines our \textbf{Coupled Author-Paper Scoring} (\textbf{CAPS}) algorithm:
\begin{align}
	x^{(k)} &= W C^{T} W^{T} x^{(k-1)} \label{eqn:rule-1} \\
	y^{(k)} &= C^{T} W^{T} x^{(k)} \label{eqn:rule-2}
\end{align}
Following previous conventions~\citep{kleinberg1999authoritative,2015arXiv150104894P}, we start with an initial guess vector (specifically, $x^{(0)} = \mathbb{1}_{m \times 1}$ and $y^{(0)} = \mathbb{1}_{n \times 1}$) and determine the values of scores iteratively (i.e.~iterate $k \geq 1$ until convergence is achieved for a given tolerance level). 

Equation~\ref{eqn:rule-1} quantifies the criterion that ``\emph{a good author is cited by good authors}''.
Equation~\ref{eqn:rule-2} quantifies the criterion that ``\emph{a good paper is cited by good authors}''. 
The equations above provide a self-consistent basis for repeated improvement~\citep[pp. 355--356]{easley2010networks}.
This can be seen by writing $L = W C$:
\begin{align}
	x^{(k)} &= W L^{T} x^{(k-1)} = W y^{(k-1)}\label{eqn:rule-1-simplified} \\
	y^{(k)} &= L^{T} x^{(k)} \label{eqn:rule-2-simplified}
\end{align}
Hence, \emph{a good author has good papers that are cited by good authors who have good papers} and so on.
The $m \times n$ matrix $L$ has entries $(L)_{ij} = \sum_{p=1}^n W_{ip} C_{pj}$ which correspond to the cumulative fractional citations made by citing author $i$ through papers $p$ (if authored by $i$) to some cited paper $j$.
Essentially, $L$ encodes the \emph{author-paper citation matrix}.

Entries of the $m \times m$ matrix product $WL^T$ in Equation~\ref{eqn:rule-1-simplified} corresponds to the cumulative fractional citations received by authors in row $i$ from authors in column $a$.
This is because $(W L^T)_{ia} = \sum_{p_1,p_2=1}^n W_{ap_2} C_{p_2 p_1} W_{i p_1}$ signifies that author $a$ in paper $p_2$ cites paper $p_1$ which is (co)authored by $i$.
The sum over all possible papers $p_1$ serves to aggregate all fractional citations received by author $i$ from author $j$.
$WL^T$ is thus the (fractional) \emph{author citation matrix}.

In effect, we find that the author score defined in Equation~\ref{eqn:rule-1-simplified} therefore corresponds to $x_i^{(k)} = \sum_{a=1}^m \sum_{p_1,p_2=1}^n W_{ap_2} C_{p_2 p_1} \allowbreak W_{ip_1} x^{(k-1)}_a$.
Therefore, the author score for author $i$ is proportional to the cumulative author citations received as well as the score of the citing authors.
This captures the intuition that \emph{authors promote each other through their published works}.
Similarly, Equation~\ref{eqn:rule-2-simplified} implies that the paper score for paper $j$ is $y^{(k)}_j = \sum_{i=1}^m L_{ij} x^{(k)}_i$.
This quantifies the relationship that \emph{the relative importance of a paper depends on the authority its citing authors}.

\section{Empirical test}
\label{sec:caps-data}
We test the CITEX and CAPS algorithm on papers published under the Thomson ISI \emph{Journal Citation Reports} (JCR) subject category of ``\emph{Information Science \& Library Science}'' (LIS) from the years 1980 up to 2012 inclusive.
This dataset consists of 213,530 papers, 471,191 total inter-paper citations, and 73,597 author keywords.
We do not conduct author or bibliographic reference disambiguation in order to assess the output quality of CAPS and CITEX when used with minimal data preprocessing.

\subsection{Authors}
\label{sec:analysis-authors}
The output of a ranking scheme depends on how it scores selected features that are present (or absent) for each datum relative to the rest of the dataset.
In general, it is difficult to determine the performance of the underlying scoring algorithm when there is no ground truth to base such judgements.
In cases like this, the most sensible thing to do is to speak of the properties of the scores generated by the algorithm of interest, and whether the rankings generated show reasonable agreement with known methods and observations.

In this respect, the distribution of author scores for CAPS and CITEX exhibit a reasonably high Spearman rank correlation coefficient ($\rho$) with $h$-index score ($p < 0.01)$: specifically, 0.77 and 0.69 for CAPS and CITEX, respectively. 
The $h$-index~\citep{hirsch2005index} provides a useful comparison to CAPS and CITEX as it too combines publication and citation traits together.
However, unlike CAPS (and to a lesser extent, CITEX), the $h$-index is not designed to differentiate whether a citation is received from a relatively ``good'' paper (author) or otherwise, hence some disparity in the resulting ranking is to be expected.
This can be seen in Table~\ref{tbl:caps-top25-authors}.

\begin{table}[hbt!]
\caption{Top 25 (out of 73,597) authors by publication count, citation count, CAPS author score, CITEX author score, and $h$-index, respectively.
Note that $h$-index values in columns denoted by $h$ are computed using available data (only ISI papers indexed under LIS JCR category from 1980-2012).
Note the usage of ordinal ranking for the $h$-index column.
}
\label{tbl:caps-top25-authors}
{\scriptsize
\begin{flushleft}
\begin{tabular}{|l|ll|ll|ll|ll|ll|}
\hline
 & \multicolumn{2}{c|}{Pubs.} & \multicolumn{2}{c|}{Times Cited} & \multicolumn{2}{c|}{CAPS} & \multicolumn{2}{c|}{CITEX} & \multicolumn{2}{c|}{$h$-index} \tabularnewline
Rank & $h$ & Author key & $h$ & Author key & $h$ & Author key & $h$ & Author key & $h$ & Author key\tabularnewline
\hline
1 & 5 & rogers.m & 5 & davis.fd & 21 & egghe.l & 5 & rogers.m & 30 & glanzel.w\tabularnewline
2 & 0 & cassada.j & 29 & benbasat.i & 26 & leydesdorff.l & 0 & cassada.j & 29 & bates.dw\tabularnewline
3 & 0 & klett.re & 12 & venkatesh.v & 23 & rousseau.r & 0 & klett.re & 29 & benbasat.i\tabularnewline
4 & 1 & ramsdell.k & 29 & bates.dw & 30 & glanzel.w & 1 & ramsdell.k & 28 & garfield.e\tabularnewline
5 & 1 & christian.g & 1 & pawlak.z & 24 & thelwall.m & 1 & christian.g & 26 & leydesdorff.l\tabularnewline
6 & 0 & vicarel.ja & 19 & straub.dw & 16 & burrell.ql & 0 & vicarel.ja & 25 & schubert.a\tabularnewline
7 & 2 & hoffert.b & 30 & glanzel.w & 25 & schubert.a & 2 & hoffert.b & 25 & spink.a\tabularnewline
8 & 1 & sutton.j & 1 & gruber.tr & 17 & ingwersen.p & 1 & sutton.j & 24 & grover.v\tabularnewline
9 & 1 & sutton.jc & 25 & spink.a & 17 & bar-ilan.j & 24 & thelwall.m & 24 & moed.hf\tabularnewline
10 & 1 & bigelow.d & 14 & salton.g & 22 & braun.t & 21 & egghe.l & 24 & thelwall.m\tabularnewline
11 & 1 & stevens.n & 3 & furnas.gw & 18 & cronin.b & 30 & glanzel.w & 23 & rousseau.r\tabularnewline
12 & 0 & zlendich.j & 24 & grover.v & 17 & van.raan.afj & 26 & leydesdorff.l & 22 & braun.t\tabularnewline
13 & 0 & fairchild.ca & 3 & deerwester.s & 16 & white.hd & 1 & sutton.jc & 21 & egghe.l\tabularnewline
14 & 1 & pearl.n & 3 & dumais.st & 12 & jacso.p & 2 & decandido.ga & 20 & willett.p\tabularnewline
15 & 0 & richard.o & 2 & landauer.tk & 24 & moed.hf & 23 & rousseau.r & 19 & ford.n\tabularnewline
16 & 2 & gordon.rs & 8 & buckley.c & 15 & vinkler.p & 3 & stlifer.e & 19 & saracevic.t\tabularnewline
17 & 0 & maccann.d & 25 & schubert.a & 16 & small.h & 1 & bigelow.d & 19 & smaglik.p\tabularnewline
18 & 0 & lombardo.d & 5 & morris.mg & 18 & mccain.kw & 25 & schubert.a & 19 & straub.dw\tabularnewline
19 & 1 & williamson.ga & 1 & harshman.r & 16 & bornmann.l & 2 & rawlinson.n & 18 & bates.mj\tabularnewline
20 & 5 & butler.t & 7 & todd.pa & 28 & garfield.e & 0 & davidson.a & 18 & chen.hc\tabularnewline
21 & 1 & raiteri.s & 9 & karahanna.e & 9 & pao.ml & 0 & de.baron.fhk & 18 & cronin.b\tabularnewline
22 & 1 & gillespie.t & 26 & leydesdorff.l & 15 & vaughan.l & 0 & elizabeth.p & 18 & dennis.ar\tabularnewline
23 & 1 & campbell.p & 18 & zmud.rw & 6 & rao.ikr & 1 & furlong.cw & 18 & lyytinen.k\tabularnewline
24 & 3 & burns.a & 14 & gefen.d & 15 & daniel.hd & 0 & hammett.d & 18 & mccain.kw\tabularnewline
25 & 2 & wyatt.n & 19 & saracevic.t & 15 & oppenheim.c & 0 & hemingway.h & 18 & zmud.rw\tabularnewline
\hline
\end{tabular}
\par\end{flushleft}
}
\end{table}

Since CAPS and CITEX are also positively correlated with $\rho = 0.85$ ($p < 0.01$), we can expect that the $h$-index distribution for top $N$ ranks by CAPS and CITEX score to resemble each other for increasingly large $N$.
For the top $N=25$ ranks, $\mu_{\mathrm{CAPS}}(h) = 18.44$ while $\mu_{\mathrm{CITEX}}(h)=6.76$.
For $N=250$ the mean $h$-index values are 8.03 and 7.03, 
while for $N=2500$ we obtain 3.32 and 3.36 for CAPS and CITEX, respectively. 
Ideally, the top percentile of any ranking should correspond to an easily interpreted ordering by quality, hence in this sense, CAPS improves on the CITEX author ranking (since the top ranks tend to correspond to high $h$-index values).

Incidentally, the top ranked author by CITEX (Rogers, with a score of $3.37 \times 10^{-5}$) corresponds to 83.6\% of the entire CITEX author score distribution.
Together with Cassada ($\mathrm{author~score}=2.39 \times 10^{-14}$), both authors take up a shocking 96\% of total scores.
Over the entire list of authors, this corresponds to a Gini coefficient\footnote{The Gini coefficient is a measure of statistical dispersion typically used to measure the level of inequality in a given sample.
For a sample of size $n$ ordered such that $x_i \leq x_{i+1}$, it is given by $G = \frac{2 \sum_{i=1}^n i x_i}{n \sum_{i=1}^n x_i} - \frac{n+1}{n}$.
A Gini coefficient of 1 indicates maximal inequality whereby the total score is associated to only one element in the sample while the remainder of the sample contributes nothing to the total score.
A Gini coefficient of 0 indicates perfect equality whereby the total score is distributed equally among all elements in the sample.
} 
of 0.9999. 
In contrast, 20\% (14,719) of top scoring authors according to the CAPS algorithm accounts for approximately 99.96\% of the scores (corresponds to a Gini coefficient of 0.9891). 
This implies that the difference between CAPS author scores for adjacent ranks becomes progressively smaller as we go down the ranks.
This is exaggerated to a greater extreme in CITEX.

Interestingly, the Gini coefficients for fractional publication count and fractional citation count of authors in the LIS dataset are 0.7744 and 0.8715, respectively.
Furthermore, 20\% of top authors account for 81.4\% of the total fractional publications  as well as 90\% of the total fractional citations.
While these values are characteristic of high levels of inequailty, they are quite tame compared to the level of inequality implied by CAPS.
The presence of such extreme levels of inequality suggests a vast differential in the ability of LIS researchers to capitalize the resources, technical skills, and opportunities at their disposal~\citep{shockley1957statistics}.

\subsection{Papers}
\label{sec:analysis-papers}
The top 25 ranking by citation count, CAPS paper score, and CITEX paper score is as displayed in Table~\ref{tbl:caps-top25-papers}.
The topmost ranks of CITEX are populated by papers sharing the same high-scoring author (Rogers).
Looking beyond the top 25 ranks, we find that with the exception of papers at ranks 7 to 12, the first 3819 positions are papers authored by Rogers, while the next 2610 positions (ranks $3820 - 6429$) are papers authored by Cassada.
Hence, CITEX tends to over-represent the importance of papers from the same highly scored author even if these do not correspond to ``high impact'' works or works that impact ``high impact works''.
This is precisely the effect we described in Section~\ref{sec:citex-algo-inspection}.

As we have seen in the case of authors, the paper citation data shows high inequality since the top 10\% of cited papers accounts for nearly 88.8\% of total citations.
This is expected since only a fraction of papers are cited and each of these papers receives additional citation in-links at a rate proportional to their current number of citation in-links.
This suggests that the citation distribution is governed by a cumulative advantage/preferential attachment process whereby the \emph{rich get richer}~\citep{price1976general,barabasi1999mean}.

\begin{table}[hbt!]
\caption{Top 25 (out of 213,530) papers by citation count, CAPS paper score, and CITEX paper score.
Papers are identified by publication year, followed by source journal abbreviation, volume, page number, and first author.
Source journal abbreviations are listed in Table~\ref{tbl:caps-journal-composition}.
TC designates the times cited for papers as reported by ISI in 2012.
The Spearman rank correlation coefficients ($p < 0.01$) over all papers are: $\rho(C_1,C_2) = 0.87$, $\rho(C_1,C_3) = 0.17$, and $\rho(C_2,C_3)=0.26$.
CAPS appears in better agreement with citation count than CITEX.
}
\label{tbl:caps-top25-papers}
{\scriptsize
\begin{flushleft}
\begin{tabular}{|l|ll|ll|ll|}
\hline
 & \multicolumn{2}{c|}{Citation count ($C_1$)} & \multicolumn{2}{c|}{CAPS ($C_2$)} & \multicolumn{2}{c|}{CITEX ($C_3$)}\tabularnewline
Rank & Paper & TC & Paper & TC & Paper & TC\tabularnewline
\hline
1 & 1982/IJCIS/11/341/pawlak & 3319 & 2006/SCI/69/121/egghe & 105 & 1995/LJ/120/113/rogers & 1\tabularnewline
2 & 1989/MISQ/13/319/davis & 3251 & 1990/JIS/16/17/egghe & 61 & 1995/LJ/120/119/rogers & 1\tabularnewline
3 & 1993/KA/5/199/gruber & 2618 & 2006/SCI/69/131/egghe & 250 & 1995/LJ/120/130/rogers & 1\tabularnewline
4 & 1990/JASIS/41/391/deerwester & 2150 & 1998/JD/54/236/ingwersen & 199 & 1995/LJ/120/187/rogers & 1\tabularnewline
5 & 1980/PAL/14/130/porter & 1653 & 2005/S/19/8/braun & 86 & 1995/LJ/120/213/rogers & 1\tabularnewline
6 & 2003/MISQ/27/425/venkatesh & 1534 & 2003/JASIST/54/550/ahlgren & 123 & 1996/LJ/121/100/rogers & 1\tabularnewline
7 & 1988/IPM/24/513/salton & 1449 & 2006/SCI/69/169/braun & 127 & 2011/LJ/136/30/fox & 0\tabularnewline
8 & 2001/MISQ/25/107/alavi & 1075 & 2006/SCI/67/491/van.raan & 177 & 2007/LJ/132/36/albanese & 4\tabularnewline
9 & 1995/ISR/6/144/taylor & 1021 & 1999/JD/55/577/smith & 93 & 1993/LJ/118/32/berry & 2\tabularnewline
10 & 2003/JMIS/19/9/delone & 772 & 2001/JASIST/52/1157/thelwall & 94 & 1989/LJ/114/18/decandido & 1\tabularnewline
11 & 2004/MISQ/28/75/hevner & 724 & 1985/JD/41/173/egghe & 48 & 1989/LJ/114/57/decandido & 0\tabularnewline
12 & 1995/MISQ/19/189/compeau & 684 & 1992/IPM/28/201/egghe & 41 & 1995/LJ/120/12/stlifer & 1\tabularnewline
13 & 2003/MISQ/27/51/gefen & 677 & 1989/SCI/16/3/schubert & 165 & 1992/LJ/117/52/rogers & 0\tabularnewline
14 & 2000/ISR/11/342/venkatesh & 596 & 1997/JD/53/404/almind & 163 & 2000/LJ/125/91/rogers & 0\tabularnewline
15 & 1999/MISQ/23/67/klein & 569 & 2001/SCI/50/65/bjorneborn & 93 & 2005/LJ/130/172/rogers & 0\tabularnewline
16 & 2000/MISQ/24/169/bharadwaj & 568 & 1986/SCI/9/281/schubert & 162 & 2006/LJ/131/114/rogers & 0\tabularnewline
17 & 1992/MISQ/16/227/adams & 542 & 2006/SCI/67/315/glanzel & 88 & 2006/LJ/131/114/rogers & 0\tabularnewline
18 & 1995/MISQ/19/213/goodhue & 540 & 2006/SCI/69/161/banks & 60 & 2006/LJ/131/123/rogers & 0\tabularnewline
19 & 1987/MISQ/11/369/benbasat & 526 & 1996/SCI/36/97/egghe & 31 & 2006/LJ/131/123/rogers & 0\tabularnewline
20 & 1999/MISQ/23/183/karahanna & 513 & 1991/JASIS/42/479/egghe & 29 & 2007/LJ/132/132/rogers & 1\tabularnewline
21 & 1999/JAMIA/6/313/bates & 497 & 2003/SCI/56/357/glanzel & 82 & 2007/LJ/132/132/rogers & 1\tabularnewline
22 & 1988/MISQ/12/259/doll & 477 & 1986/SCI/9/103/leydesdorff & 46 & 2007/LJ/132/171/rogers & 0\tabularnewline
23 & 1999/IJGIS/13/143/stockwell & 475 & 2001/SCI/50/7/bar-ilan & 64 & 2007/LJ/132/96/rogers & 0\tabularnewline
24 & 2000/MISQ/24/115/venkatesh & 475 & 2002/JASIST/53/995/thelwall & 72 & 2006/LJ/131/27/rogers & 1\tabularnewline
25 & 2003/ISR/14/189/chin & 472 & 1996/JIS/22/165/egghe & 24 & 2003/LJ/128/40/rogers & 2\tabularnewline
\hline
\end{tabular}
\par\end{flushleft}
}
\end{table}

\begin{table}[hbt!]
\caption{Journal composition in top 100 ranks by algorithm.}
\label{tbl:caps-journal-composition}
{\scriptsize
\begin{tabular}{|llr|llr|llr|}
\hline 
\multicolumn{3}{|c|}{Citation count} & \multicolumn{3}{c|}{CAPS} & \multicolumn{3}{c|}{CITEX}\tabularnewline
\hline 
MISQ & mis.quart & 43 & SCI & scientometrics & 33 & LJ & libr.j & 100 \tabularnewline
ISR & inform.syst.res & 14 & JASIS & j.am.soc.inform.sci & 17 & &  & \tabularnewline
JAMIA & j.am.med.inform.assn & 9 & JASIST & j.am.soc.inf.sci.tec & 14 & &  & \tabularnewline
JASIS & j.am.soc.inform.sci & 6 & JD & j.doc & 12 & &  & \tabularnewline
JD &  j.doc & 3 & JIS & j.inform.sci & 9 & &  & \tabularnewline
IPM & inform.process.manag & 3 & IPM & inform.process.manag & 6 & &  & \tabularnewline
JMIS & j.manage.inform.syst & 3 & JI & j.informetr & 5 & &  & \tabularnewline
IJCIS & int.j.comput.inf.sci & 2 & ARIS & annu.rev.inform.sci & 2 & &  & \tabularnewline
IM & inform.manage & 2 & SSI & soc.sci.inform & 1 & &  & \tabularnewline
IJGIS & int.j.geogr.inf.sci & 2 & S & scientist & 1 & &  & \tabularnewline
SCI & scientometrics & 2 & & &  &  &  & \tabularnewline
ARIS & annu.rev.inform.sci & 1 & & &  &  &  & \tabularnewline
CJIS & can.j.inform.sci & 1 & & &  &  &  & \tabularnewline
EJIS & eur.j.inform.syst & 1 & & &  &  &  & \tabularnewline
GIQ & gov.inform.q & 1 & & &  &  &  & \tabularnewline
IJGIS & int.j.geogr.inf.syst & 1 & & &  &  &  & \tabularnewline
IMA & inform.manage-amster & 1 & & &  &  &  & \tabularnewline
JASIST & j.am.soc.inf.sci.tec & 1 & & &  &  &  & \tabularnewline
JIS & j.inf.sci & 1 & & &  &  &  & \tabularnewline
KA & knowl.acquis & 1 & & &  &  &  & \tabularnewline
OR & online.rev & 1 & & &  &  &  & \tabularnewline
PAL & program-autom.libr & 1 & & &  &  &  & \tabularnewline
\hline 
\end{tabular}
}
\end{table}

In contrast, the CAPS paper score possesses a Gini coefficient of 0.9912, while CITEX has a slightly lower value of 0.9785.
This implies that both methods exhibit large score differentials only between the topmost ranks.
For CAPS paper score, this can be traced to the fact that 81.2\% of the lowest scoring population has a score of exactly zero (76\% of papers in the study data have zero citations\footnote{The LIS dataset consists of 103,768 papers from \emph{Library Journal} ($\sim 48.6$\% of total).
This is nearly 14 times larger than the the $2^{nd}$ largest contributor, i.e.~\emph{Scientist}.
While this seems excessively high, consider that only 1.9\% of papers from \emph{Library Journal} contributes 1\% of non-zero citations in the LIS dataset (from a total of 471,191 citations for 213,530 papers).
In comparison, \emph{Scientometrics} is only the $6^{th}$ largest contributor to the dataset with 3100 papers (1.5\% of total LIS papers) yet contributes a total of 29792 citations (6.3\% from LIS total) making it the $4^{th}$ largest contributor citation-wise. 
For reference, the largest citation counts are attributed to \emph{MIS Quarterly}, \emph{J AM MED INFORM ASSN}, and \emph{J AM SOC INFORM SCI} with 55736, 30470, 30317 citations, respectively.
}).
The reason for this is that the coupling of both author features and paper features places strict limits on the size of the non-zero scoring population. 
On the other hand, the CITEX paper score has no zero scoring population (due to the presence of artificial paper self-citations).
The extremely high Gini coefficients for both CITEX and CAPS\footnote{For CITEX, 1\% of the top scoring population accounts for 50.1\% of the total score, while 2\% accounts for 91\%.
In comparison, CAPS has 1\% and 2\% of the top scoring population accounting for 88\% and 97\% of total scores, respectively.
}
implies that we can only reasonably differentiate a small fraction of the dataset corresponding to top scoring papers that coincide with top scoring authors.

A quick glance at top scoring papers listed in the ``Citation count'' column of Table~\ref{tbl:caps-top25-papers} reveals that these mostly correspond to informatics papers rather than informetrics.
Contrast this with the listing shown in the ``CAPS'' column where the emphasis is more towards informetrics papers instead.
The reason for this is that the CAPS algorithm takes into account authorship features when scoring papers, which are not accounted for in a simple citation count.
Since informetrics authors are highlighted in Table~\ref{tbl:caps-top25-authors}, it follows that informetrics papers are also highlighted in Table~\ref{tbl:caps-top25-papers}.
Table~\ref{tbl:caps-journal-composition} provides a listing of journals in the top 100 ranks.
This provides some indication of the research field predominantly featured by each method.

\section{Conclusion}
\label{sec:caps-conclusion}
In this paper we have constructed a modified version of the CITEX algorithm originally introduced by~\cite{2015arXiv150104894P}.
This algorithm was designed to assign relative importance scores to papers and authors by taking into account data from both entities simultaneously.
Conventional methods like citation count and PageRank, for example, cannot do so without appropriate modification.
The modification of CITEX which we propose, dubbed the CAPS (Coupled Author-Paper Scoring) algorithm, is designed to address some of the weaknesses of Pal and Ruj's original algorithm which we described in Section~\ref{sec:citex-algo-inspection} (essentially, the shortcomings can be traced to artificially introduced self-citations on the paper-level).

Using a real dataset (ISI papers published from $1980-2012$ in the JCR subject category of ``\emph{Information Science \& Library Science}''), we show that our proposed modifications outperforms CITEX in identifying important authors and papers.
However, the CAPS algorithm appears to suffer from high inequality in the resulting score distributions as indicated by an extremely high Gini coefficient ($\sim 0.99$).
The inequality is similarly pronounced in CITEX.
This implies that both CAPS and CITEX generate extreme prejudice in the allocation of scores to the top scoring minority.

However, this is not necessarily a bad thing.
By design, CAPS allocates high scores to authors and papers associated to instances where the likelihood of future success (an increase in publication count or citation count) is proportional to previous success.
Hence, CAPS can be used to highlight parts of the data attributed to the ``\emph{rich get richer}'' effect.
In contrast, CITEX rewards high scores for authors lying at the tail of the publication productivity distribution, and by association, rewards high scores for papers published by such authors irrespective of the relative importance of their papers within the paper citation network.
In this sense, CITEX is useful to find instances where high productivity is mismatched with low impact.

While bibliometric analytic algorithms such as CITEX or CAPS, or even bibliometric adaptations of website ranking algorithms such as HITS or PageRank can prove useful in identifying what is important in a given dataset, it is crucial to be aware of the limitations and subtleties of such methods. 
Each method finds exactly what it is designed to seek and since it is hard to account for, let alone anticipate every relevant feature or contingency, we must concede that the rankings produced are themselves only facets of the underlying organization in the data.
Hence, bibliometric analytic algorithms should be used first and foremost to guide decisions on where to look deeper (i.e.~to construct recommendation engines), and if necessary, used with extreme caution when drawing inferences on the relative standing of bibliometric entities.

\bibliographystyle{apalike}
\bibliography{refs}   

\end{document}